\documentclass[12pt,a4paper]{article}
\usepackage{xcolor,paralist}
\usepackage{natbib,soul}
\usepackage[pdftex,colorlinks=true,hypertexnames=false]{hyperref}
\definecolor{darkblue}{rgb}{0,0,.6}
\hypersetup{citecolor=darkblue,linkcolor=darkblue,urlcolor=darkblue}
\usepackage{amssymb,enumitem,booktabs,subfig,setspace,bm,longtable,xr,dsfont,orcidlink}
\usepackage[top=1in, bottom=1in, left=1.1in, right=1.1in]{geometry}
\usepackage{url}
\usepackage{amsmath}
\usepackage{amsfonts}
\usepackage{epsfig}
\usepackage{graphics}
\usepackage[normalem]{ulem}
\usepackage{palatino,eulervm}
\usepackage{graphicx, soul}
\usepackage{lscape}
\usepackage{rotating}
\usepackage[final,nopatch=footnote]{microtype}
\usepackage[linewidth=1pt]{mdframed}
\allowdisplaybreaks[4]

\providecommand{\U}[1]{\protect\rule{.1in}{.1in}}

\newsavebox\CBox

\usepackage{amsthm,thmtools}
\usepackage{mathrsfs}

\makeatletter
\def\th@newremark{\th@remark\thm@headfont{\bfseries}}
\makeatletter
\theoremstyle{newremark}

\declaretheoremstyle[spaceabove=6pt, spacebelow=6pt, headfont=\bfseries, notefont=\mdseries, notebraces={(}{)}, bodyfont=\normalfont, postheadspace=0.5em]{mystyle}

\makeatletter
\newcommand*{\addFileDependency}[1]{
\typeout{(#1)}
\@addtofilelist{#1}
\IfFileExists{#1}{}{\typeout{No file #1.}}
}\makeatother

\newcommand{\Rlogo}{\protect\includegraphics[height=1.8ex,keepaspectratio]{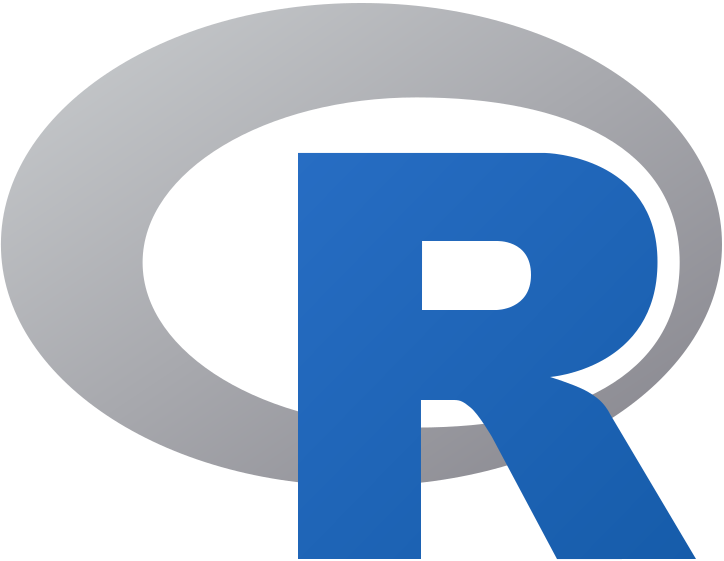}}

\begin{document}

\title{Age-period modeling of mortality gaps: the cases of cancer and circulatory diseases}

\author{\normalsize Giacomo Lanfiuti Baldi \\
\normalsize Department of Statistics \\ 
\normalsize Sapienza University of Rome \\
\\
\normalsize Andrea Nigri \orcidlink{0000-0002-2707-3678} \\
\normalsize Department of Economics, Management and Territory \\
\normalsize University of Foggia \\
\\
\normalsize Han Lin Shang \orcidlink{0000-0003-1769-6430}  \\
\normalsize Department of Actuarial Studies and Business Analytics \\
\normalsize Macquarie University
}

\date{}

\maketitle

\centerline{\bf Abstract}

\medskip

Understanding and modeling mortality patterns, especially differences in mortality rates between populations, is vital for demographic analysis and public health planning. We compare three statistical models within the age-period framework to examine differences in death counts. The models are based on the double Poisson, bivariate Poisson, and Skellam distributions, each of which provides unique strengths in capturing underlying mortality trends. Focusing on mortality data from 1960 to 2015, we analyze the two leading causes of death in Italy, which exhibit significant temporal and age-related variations. Our results reveal that the Skellam distribution offers superior accuracy and simplicity in capturing mortality differentials. These findings highlight the potential of the Skellam distribution for analyzing mortality gaps effectively.

\vspace{-.15in}

\begin{flushleft}
\textbf{Keywords}: Italian age-period data; Mortality modeling; Causes-of-deaths; Skellam regression 
\end{flushleft}

\newpage
\doublespacing

\section{Introduction}

Since the early nineteenth century, life expectancy in developed countries has shown consistent growth, particularly since the end of the Second World War \citep{oeppen2002broken}. Researchers, governments, and related organizations have often noted a faster-than-expected increase in longevity, posing financial challenges for life insurers, pension plans, and social security systems. A shift in longevity trends can substantially affect the number of elderly individuals and centenarians, thereby increasing demand for social services. Consequently, official forecasts that guide future pensions, healthcare, and other social needs can shape individual decisions on savings and retirement timing. In this context, methodological advancements in longevity studies have become increasingly important for anticipating future changes \citep[see, e.g.,][]{torri2012forecasting, raftery2013bayesian, pascariu2018double}, aiming to reduce the gap between actual outcomes and projections.

Various approaches have been utilized to model the mortality surface, capturing how mortality rates evolve over time. Prior to the 1980s, mortality models were relatively simple and often depended on expert judgment \citep[see][for a detailed review]{Pollard87}. However, with the increased availability of mortality data and the advancement of statistical methods, more sophisticated and complex mortality models have since been developed.

According to \cite{booth2008mortality}, three main paradigms in demographic modeling can be identified. The first paradigm, \textit{explanation}, is based on structural or epidemiological models that focus on specific causes of death, such as a well-established link between lung cancer and tobacco smoking. The second paradigm, \textit{expectation}, relies on expert opinions and may range from informal assessments to more formalized approaches. The third paradigm, \textit{extrapolation}, utilizes regular patterns observed in age-specific mortality rates and temporal trends. This approach includes complex stochastic models, such as the \citeauthor{lee1992modeling}'s \citeyearpar{lee1992modeling} model and the generalized age-period-cohort model. Although the Lee–Carter model is regarded as one of the most influential in mortality modeling and forecasting, various extensions and modifications have gained considerable attention in actuarial and demographic literature \citep[see, e.g.,][]{brouhns2002poisson, haberman2011comparative, cairns2006two, cairns2009quantitative, Nigri2019, Marino2023, Shang2020, Shang2023}.

Despite the widespread use of the Lee–Carter model and its variants, which serve as benchmarks for many new methodologies, these models have several limitations. \cite{cairns2008modelling} addressed the challenge of optimal mortality estimation by proposing criteria for a robust mortality model, emphasizing consistency with historical data and biologically plausible long-term dynamics. Apart from modeling age-specific mortality rates, research has shifted toward developing statistical frameworks that model the age-at-death distribution, allowing for parameter forecasting using econometric models \citep[see, e.g.,][]{basellini2019modelling, pascariu2019maximum, aliverti2022dynamic, Cardillo2023}. However, researchers and practitioners have noted that separate forecasts for different populations or sub-populations can lead to unreliable scenarios, as backtesting often reveals diverging behaviors. Such inconsistencies in forecasts are implausible and may undermine decision-making processes.

Building on this discussion, we propose a novel approach for modeling and forecasting the gap in mortality counts between two populations. Specifically, we leverage recent developments in cause-of-death analysis \citep[see, e.g.,][]{arnold2015causes, kjaergaard2019forecasting, wilmoth1995mortality} by treating the two target populations as distinct causes of death within the same country. These two causes of death have implications for public health, particularly in identifying trends and temporal changes across different age groups. Understanding the trend associated with any specific cause of death enables policymakers to better allocate and target healthcare resources.

Our focus, therefore, is on the distribution of deaths, highlighting the disparities between two population counts, which represent the two leading causes of death, namely cancer and circulatory diseases, in the Italian population. These causes are major contributors to overall mortality. It is noteworthy that the two populations could also represent other categories, such as different genders, countries, or sub-populations within the same country. We aim to provide a more reliable and coherent forecast by addressing the inconsistencies often seen in separate forecasts for different populations or sub-populations. We develop a statistical toolkit for those working with mortality data, allowing for the joint modeling of mortality evolution between two populations while considering their differences. Alongside traditional alternatives like the double and bivariate Poisson distributions, we introduce the Skellam distribution, commonly used in fields involving count data, such as sport analysis.

Recently, \cite{KLW21} introduced the Skellam distribution within the Lee-Carter framework to model the log-difference of mortality rates, focusing on age-dependent improvements in mortality and their implications for pricing longevity derivatives. While their work provides valuable insights into the dynamics of mortality improvement, our approach differs fundamentally in both scope and methodology. Specifically, we employ the Skellam distribution within the age-period framework to model differences in mortality counts across causes of death. In this context, it is standard practice to fix the age profile parameters, as the primary focus lies on capturing temporal (period) effects. This distinction highlights the originality of our contribution, which centers on analyzing mortality gaps with a straightforward and coherent structure tailored to demographic and public health applications.

Mortality data are commonly reported as count data within specific age-period domains for various populations. The simplest models for modeling mortality patterns use age and period as the only explanatory variables, since they capture all the necessary information intrinsic to the data. This approach serves as a foundational basis for exploring and understanding the complex dynamics of mortality across different populations.

The remainder of the paper is structured as follows: Section~\ref{sec:Data and Measures} introduces the motivating dataset and demographic measures. Sections~\ref{Sec:Models} and~\ref{Sec:foreca} detail the methods used in our study. In Section~\ref{sec:res}, we present the implementation and evaluation of the distributions considered. Section~\ref{sec:conc} concludes the paper and offers some insights on potential extensions of the presented methodology.

\section{Data and Measures}\label{sec:Data and Measures}

Let $x$ represent age, $\{t_1, t_2, \ldots, T\}$ denote a set of calendar years, and $\{\text{A}, \text{B}\}$ represent the set of populations we consider. The death count at age $x$ and time $t$ for population $i$ is denoted by $D_{xt}^{i}$.

We utilize period life tables by sex and single year of age (0–85+) from the \cite{HMD}. We extract death counts by cause of death from the \cite{WHO} for the total population. This allows us to compute the proportion of deaths by cause, age group (in 5-year intervals), and sex for each year from 1960 to 2015.

The data, in their original form, were classified using the seventh, eighth, ninth, and tenth revisions of the International Classification of Diseases (ICD). To analyze factors influencing Italian longevity, we group deaths into two major categories up to age 85+ (see Table~\ref{tab:ICD}): cancers and circulatory system diseases. These categories represent significant causes of death, capturing a substantial proportion of overall mortality. For details on ICD codes and the classification system, refer to Table~\ref{tab:ICD}.
\begin{table}[!htb]
\centering
\tabcolsep 0.18in
\caption{\small{ICD codes and our classification into two major groups. \textit{Source}: authors' own elaborations on data from the \cite{WHO}.}}
\begin{tabular}{@{}lllll@{}}
\toprule
\textbf{Cause} & \textbf{ICD7} & \textbf{ICD8} & \textbf{ICD9} & \textbf{ICD10} \\ \midrule
Cancer & \begin{tabular}[c]{@{}l@{}}A044-A059, \\ B018-B019\end{tabular}                    & \begin{tabular}[c]{@{}l@{}}A045-A060\end{tabular}& \begin{tabular}[c]{@{}l@{}}B08-B09, B100, \\ B101, B109, B11-B17\end{tabular}& \begin{tabular}[c]{@{}l@{}} C \end{tabular}\\ \hline
Circulatory            & \begin{tabular}[c]{@{}l@{}}A070, A079-A086,\\ B022, B024-B029\end{tabular}  & A080-A088  & B25-B30 & I00-I99 \\
\bottomrule
\end{tabular}
\label{tab:ICD}
\end{table}

We define the gap in deaths between two populations, in our case the two causes of death $i$, at age $x$ and year $t$ as:
\begin{equation*}
G_{xt}= D^{A}_{xt}-D^{B}_{xt}.
\end{equation*}

\begin{figure}[!htb]
\centering
\includegraphics[width=17cm]{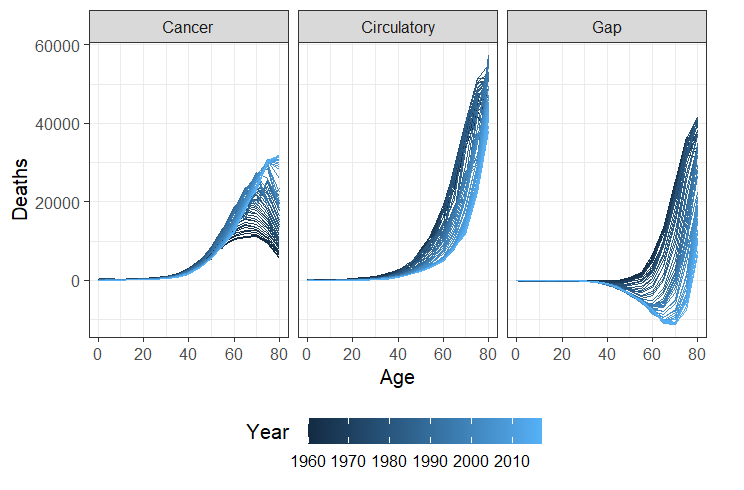}
\caption{\small{Trends in cancer and circulatory death counts, as well as their differences, by age and year in Italy (both sexes combined) from 1960 to 2018.}}
\label{fig:1}
\end{figure}

Figure~\ref{fig:1} provides a detailed depiction of the relationship between age and the temporal evolution of dynamics for a specific cause of death. It also illustrates the measure we are modeling: the difference in counts over time.

\section{Models and Distributions}\label{Sec:Models}

\subsection{Age-period model}

Statistical modeling in longevity analysis is essential for understanding its evolution and variability. The literature highlights three key components — age, period, and cohort effects — that provide valuable insights into non-observable causal mechanisms.
 
Age effect is associated with a specific population's biological and social processes, linked to its age structure rather than the period. In contrast, the period effect results from external factors that impact all age groups equally at a particular calendar time. These effects may arise from various environmental, social, or economic factors; the Spanish flu, the First World War, and COVID-19 are prime examples. Lastly, cohort effect refers to variations that stem from a group of individuals' unique experiences or exposures as they progress through time.

The model that integrates all these components is known as the age-period-cohort approach, which presents technical challenges that can complicate the interpretation and clarity of estimates, especially when working with causes of death \citep[see, e.g.,][for more details]{mason1973some}.

The demographic literature centering on longevity risks has primarily proposed age-period models. Continuing this line of research, we focus on the age-period framework. A canonical regression model, based on the ordinary least squares method, is specified as follows:
\begin{equation}
Y_{xt} = \mu + \beta_x^A + \beta_t^P + \varepsilon_{xt}, \label{eq:Classical APC}
\end{equation}
where $Y_{xt}$ represents the outcome variable to be explained, $\mu$ is the intercept, $\beta_x^A$ denotes the $x$\textsuperscript{th} age effect, $\beta_{t}^{P}$ represents the $t$\textsuperscript{th} period effect, and $\varepsilon_{xt}$ is the error term. In both age-period and age-period-cohort models, age and period are treated as categorical variables to capture nonlinear effects, utilizing a generalized linear model approach.

Given the mortality data for two populations, A and B, specific to ages and years, our goal is to determine the average period trend and the age-specific profile of the differences in mortality. We leverage the fundamental assumption in mortality modeling that death counts for each population, $\left( D^A_{xt}, D^B_{xt} \right)$, are generated by a Poisson process \citep{brillinger1986biometrics, basellini2019modelling} with distinct intensity parameters $\lambda^A_{xt}$ and $\lambda^B_{xt}$:
\begin{equation*}
D^{A}_{xt} \sim \text{Pois}(\lambda^A_{xt}),\qquad D^{B}_{xt} \sim \text{Pois}(\lambda^B_{xt}).
\end{equation*}

\subsection{Poisson distribution}

\subsubsection{Double Poisson}

The simplest model employs two independent Poisson variables, where the parameters are constructed as the sum of age effects ($\beta^\alpha_x$) and period effects ($\beta^\pi_t$). We refer to this model as the \textit{double Poisson}. It involves two distinct models for the parameters of the two Poisson processes:
\begin{align*}
\log(\lambda^A_{xt}) &= \mu^A + \beta^{A,\alpha}_x +\beta^{A,\pi}_{t} \\
\log(\lambda^B_{xt}) &= \mu^B + \beta^{B,\alpha}_x +\beta^{B,\pi}_{t} 
\end{align*}

If the goal is to estimate the differences in deaths between the two populations by year and age, one can calculate:
\begin{equation*}
\widehat{G}_{xt}= \widehat{\lambda}^{A}_{xt}-\widehat{\lambda}^{B}_{xt},
\end{equation*}
where $\widehat{\lambda}^{A}_{xt}$ and $\widehat{\lambda}^{B}_{xt}$ are the parameter estimates.

The joint probability of the observations \( (D_{xt}^A, D_{xt}^B) \) is:
\[
f(D_{xt}^A, D_{xt}^B;\lambda_{xt}^A, \lambda_{xt}^B ) = \frac{e^{-\lambda_{xt}^A} (\lambda_{xt}^A)^{D_{xt}^A}}{D_{xt}^A!} \cdot
\frac{e^{-\lambda_{xt}^B} (\lambda_{xt}^B)^{D_{xt}^B}}{D_{xt}^B!}.
\]

Given observations $(D_{xt}^A, D_{xt}^B)$ for all age and period combinations \((x, t)\), the log-likelihood function denoted by $\ell$ is:
\begin{equation*}
\ell = \sum_{x,t} \left[
-D_{xt}^A \log(D_{xt}^A!) - \lambda_{xt}^A + D_{xt}^A \log(\lambda_{xt}^A)
- D_{xt}^B \log(D_{xt}^B!) - \lambda_{xt}^B + D_{xt}^B \log(\lambda_{xt}^B)
\right].
\end{equation*}

\subsubsection{Bivariate Poisson}

The double Poisson model, while straightforward, makes a somewhat unrealistic assumption by treating the processes as generated by two independent Poisson distributions. In cases where we consider two subpopulations within a single population (such as two sexes or two causes of death) or two populations experiencing similar environmental and external factors (such as neighboring countries), the correlation between deaths in the two groups can be quite high.

An alternative to the double Poisson model is the \textit{bivariate Poisson} distribution, which accounts for the dependence between the two outcome variables. In this model, while the marginal distributions remain Poisson, the random variables are allowed to be dependent \citep{karlis2005bivariate}.

Consider random variables $X_k$, $k=1,2,3,$ which follow independent Poisson distributions with parameters $\lambda_k >0$. Thus, the random variables $X = X_1 + X_3$ and $Y = X_2 + X_3$ follow jointly a bivariate Poisson distribution:
\begin{equation*}
(X,Y) \sim \text{BP}(\lambda_1, \lambda_2, \lambda_3)
\end{equation*}
        
Marginally, each random variable follows a Poisson distribution: 
\begin{align*}
    X \sim \text{Pois}(\lambda_1 + \lambda_3),\quad Y \sim \text{Pois}(\lambda_2 + \lambda_3)
\end{align*}

Moreover, $\text{Cov}(X,Y) = \lambda_3$, indicating that $\lambda_3$ measures the dependence between the two random variables. If $\lambda_3 = 0$, the bivariate Poisson distribution reduces to the double Poisson model.

In the case of the death count of two populations, we have:
\begin{equation*}
(D^A_{xt},D^B_{xt}) \sim \text{BP}\left(\lambda^A_{xt}, \lambda^B_{xt}, \lambda^{AB}_{xt}\right)
\end{equation*}

Therefore, the joint probability mass function is given by:
\begin{equation*}
f(D_{xt}^A, D_{xt}^B; \lambda_{xt}^A, \lambda_{xt}^B, \lambda_{xt}^{AB})
= e^{-(\lambda_{xt}^A + \lambda_{xt}^B + \lambda_{xt}^{AB})} 
\frac{(\lambda_{xt}^A)^{D_{xt}^A}}{D_{xt}^A}\frac{(\lambda_{xt}^B)^{D_{xt}^B}}{D_{xt}^B}
\sum_{k=0}^{\min(D_{xt}^A, D_{xt}^B)} \binom{D_{xt}^A}{k} \binom{D_{xt}^B}{k} k! \left(\frac{\lambda_{xt}^{AB}}{\lambda_{xt}^{A}\lambda_{xt}^{B}}\right)^{k}. 
\end{equation*}

Given observations \((D_{xt}^A, D_{xt}^B)\) for all age and period combinations \((x, t)\), the log-likelihood function is:

\begin{equation*}
\ell = \sum_{x,t} \left[
-(\lambda_{xt}^A + \lambda_{xt}^B + \lambda_{xt}^{AB})
+ \log \left( \sum_{k=0}^{\min(D_{xt}^A, D_{xt}^B)} \binom{D_{xt}^A}{k} \binom{D_{xt}^B}{k} k! \left(\frac{\lambda_{xt}^{AB}}{\lambda_{xt}^{A}\lambda_{xt}^{B}}\right)^k  \right)
\right].
\end{equation*}

In our modeling using the bivariate Poisson distribution, we explore three distinct models, each concentrating on a different parameter.
\begin{align*}
\log(\lambda^A_{xt}) &= \mu^A + \beta^{A,\alpha}_x +\beta^{A,\pi}_{t} \\
\log(\lambda^B_{xt}) &= \mu^B + \beta^{B,\alpha}_x +\beta^{B,\pi}_{t} \\
\log(\lambda^{AB}_{xt}) &= \mu^{AB} 
\end{align*}
Following the work of \cite{karlis2005bivariate}, we consider models with a constant $\lambda^{AB}_{xt}$ (without covariates on $\lambda^{AB}_{xt}$) for ease of interpretation.

In the estimation of the mortality gap, we have simply:
\begin{align*}
\widehat{G}_{xt} &= \widehat{\lambda}^{A}_{xt} + \widehat{\lambda}^{AB}_{xt}-\widehat{\lambda}^{B}_{xt} - \widehat{\lambda}^{AB}_{xt} \\
\widehat{G}_{xt} &= \widehat{\lambda}^A_{xt}- \widehat{\lambda}^{B}_{xt}.
\end{align*}

\subsection{Skellam distribution} 

By employing the bivariate Poisson distribution to estimate the gap between deaths, the correlation component is eliminated. We propose using the \textit{Skellam} (or \textit{Poisson Difference}) distribution to directly model the gap rather than jointly modeling the number of deaths for the two populations.

The Poisson Difference distribution, introduced by \cite{skellam1946frequency}, describes the difference $Z = X - Y$ between two discrete random variables $X$ and $Y$. The probability function of $Z$ is a discrete distribution defined over the set of integers $\mathbb{Z}$, encompassing both positive and negative values. This distribution applies to any pair of random variables $(X, Y)$ that can be expressed as $X = W_1 + W_3$ and $Y = W_2 + W_3$, where $W_1 \sim \text{Pois}(\lambda_1)$ and $W_2 \sim \text{Pois}(\lambda_2)$, with $W_3$ following any distribution. The joint distribution of $X$ and $Y$ forms a bivariate distribution, with correlation arising from the shared stochastic component $W_3$ \citep{karlis2009bayesian}. The mean of a Skellam distribution is given by $\mathbb{E}[Z] = \lambda_1 - \lambda_2$.

The probability mass function of $Z$ is defined as $\operatorname{Pr}(Z=z)=f\left(z ; \lambda_1, \lambda_2\right)$ with,
\begin{equation}
f\left(z ; \lambda_1, \lambda_2\right)=e^{-\lambda_1-\lambda_2}\left(\frac{\lambda_1}{\lambda_2}\right)^{\frac{z}{2}} I_z\left(2 \sqrt{\lambda_1 \lambda_2}\right),
\end{equation}
where $I_z(v)=\left(\frac{v}{2}\right)^z \sum_{k=0}^{\infty} \frac{\left(\frac{v^2}{4}\right)^k}{k!\cdot(z+k)!}$ 
is the modified Bessel function of the first kind with order $z$ \citep[see][]{abramowitz1968handbook}.

Therefore, we assume that $G_{xt} \sim \text{PD}(\lambda^C_{xt}, \lambda^D_{xt})$, and, as before, we model it using age and period as covariates:
\begin{align*}
&\log(\lambda^C_{xt})= \mu^C + \beta^{C,\alpha}_x +\beta^{C,\pi}_{t},\\
&\log(\lambda^D_{xt})= \mu^D + \beta^{D,\alpha}_x +\beta^{D,\pi}_{t}, 
\end{align*}

We use the superscripts $^{\text{C}}$ and $^{\text{D}}$ to denote the two parameters because Skellam's parameters $\lambda^C_{xt}$ and $\lambda^D_{xt}$ do not represent the intensities of deaths for populations A and B directly. Instead, they represent two intensities that together approximate the differences in deaths.

The fitted values from the regression directly provide estimates of the death gap:
\begin{equation*}
\widehat{G}_{xt} = \widehat{\lambda}^{C}_{xt} -\widehat{\lambda}^{D}_{xt}.
\end{equation*}

The Skellam distribution requires only the series of differences in deaths between populations~A and~B, rather than the two death series individually.

We estimate the model parameters by maximizing the log-likelihood function of Skellam distribution, which is
\begin{align*}
\ell\left(\mu, \beta_x, \beta_t ; G_{xt}\right) &= \sum_{x,t} \left[ 
-(\lambda^C_{xt} + \lambda^D_{xt}) 
+ \frac{G_{xt}}{2} \log\left(\frac{\lambda^C_{xt}}{\lambda^D_{xt}}\right) 
+ \log\left(I_{G_{xt}}\left(2\sqrt{\lambda^C_{xt}\lambda^D_{xt}}\right)\right)
\right].
\end{align*}

\subsection{Estimation and goodness of fit}

With regard to the estimation procedure, \cite{HoSinger2001} proposes a generalized least squares method, whereas \cite{Kocherlakota2001} outlines a Newton–Raphson approach for likelihood maximization in the context of bivariate Poisson regression models. In this application we leverage the approach by \cite{karlis2003analysis} using the EM algorithm to obtain maximum likelihood estimates for both, double and bivariate Poisson. The EM algorithm proceeds by estimating the unobserved data via their conditional expectations at the E-step and then it maximizes the complete-data likelihood at the M-step \citep{karlis2005bivariate}.

The estimation of the parameters for the Skellam regression is performed by minimizing the negative log-likelihood using the Broyden–Fletcher–Goldfarb–Shanno (BFGS) algorithm, using the \texttt{optim} function in \Rlogo.

The parameter vector \(\bm{\theta} = \{\mu^C, \mu^D, \beta_x^C, \beta_t^D, \beta_x^D, \beta_t^D\}\) includes the intercepts (\(\mu^C, \mu^D\)), age-specific effects (\(\beta_x^C, \beta_x^D\)), and period-specific effects (\(\beta_t^C, \beta_t^D\)). The optimization is carried out iteratively, where at each step the parameters are updated according to the BFGS algorithm:
\begin{equation*}
\bm{\theta}^{(k+1)} = \bm{\theta}^{(k)} - H^{-1} \nabla (-\ell),
\end{equation*}
where $H^{-1}$ is the approximated inverse Hessian matrix, and \(\nabla (-\ell)\) represents the gradient of the negative log-likelihood with respect to the parameters.

The optimization terminates when the norm of the gradient satisfies a convergence criterion, such as \(\|\nabla (-\ell)\| < \epsilon\), where \(\epsilon\) is a predefined tolerance. This approach ensures accurate estimation of parameters while leveraging the full structure of the Skellam distribution.

In order to measure the goodness-of-fit in terms of the log-likelihood, Akaike information criterion (AIC), Akaike information criterion with small-sample correction ($\text{AIC}_{c}$), and Bayesian information criterion (BIC) are used. 
\begin{align*}
\text{AIC}&=2 p-2 \ell \\
\text{AIC}_{c}&=\text{AIC}+\frac{2 p^2+2 p}{n-p-1}
\end{align*}
and
\begin{equation*}
\text{BIC} = p \log n-2 \ell,
\end{equation*}
where $p$ is the number of effective parameters, and $n$ is the number of total observations.
All the results are presented in Tables~\ref{Tab:InOutAge0} and~\ref{Tab:InOutAge40} of Section~\ref{sec:res}.

\section{Forecasting}\label{Sec:foreca}

Introducing a novel forecasting model for the differences in mortality between two distinct populations is crucial for several reasons. It allows for generating coherent forecasts for both populations, ensuring consistency in predictive analysis. By leveraging the Skellam distribution, which naturally models the difference between two Poisson-distributed variables, we can more accurately capture the inherent variability and interdependence between the causes of death. This approach enhances the reliability of forecasts and provides a deeper understanding of the underlying mortality dynamics, ultimately supporting more informed public health decisions and better resource allocation to address these gaps.

Given the age and period effects estimated during the baseline period using the regression models discussed in the previous subsections, the proposed forecasting method involves fixing the observed age profile $\beta^{\alpha}_x$ and the intercept $\mu$ for the two populations. We then forecast the period effects $h$ years ahead $\beta^{\pi}_{T+h}$. This forecasting approach is applied uniformly across all models presented in Section~\ref{Sec:Models}.

We aim to predict the differences in mortality between the two populations $h$ years into the future:
\begin{equation*}
\widehat{G}_{x,T+h} = \widehat{\lambda}^A_{x,T+h}-\widehat{\lambda}^{B}_{x,T+h},
\end{equation*}
where
\begin{align*}
\widehat{\lambda}^{A}_{x, T+h} &= \exp\left(\mu^A + \beta^{A,\alpha}_x +\widehat{\beta}^{A,\pi}_{T+h}\right) \\
\widehat{\lambda}^{B}_{x, T+h} &= \exp\left(\mu^{B} + \beta^{B,\alpha}_x +\widehat{\beta}^{B,\pi}_{T+h}\right)
\end{align*}

We propose using a multivariate random walk model to project the historical period effects for the two populations. This approach is selected because it effectively captures the joint dynamics and correlations between the two causes of death over time. The multivariate random walk model assumes that future values evolve as a stochastic process with a deterministic drift and random fluctuations. This framework offers both flexibility and robustness in modeling temporal dependencies and inherent randomness in mortality data, resulting in more accurate and realistic forecasts.

We denote $\beta^{(i)}_t$ as the observed value of the period parameter (with superscript $\pi$ omitted for simplicity) for population $i \in \{A, B\}$ in year $t \in \{t_1, t_2, \ldots, T\}$. Accordingly, we define the matrix of period parameters as:
\begin{equation*}
\bm{\beta_{t}^{(i)}}=
\begin{bmatrix}
\beta_{t_1}^{(A)} & \beta_{t_2}^{(A)} & \dots & \beta_{T}^{(A),} \\
\beta_{t_1}^{(B)} & \beta_{t_2}^{(B)} & \dots & \beta_{T}^{(B)} \\
\end{bmatrix},
\end{equation*}
where the row indicates the specific population population and the column indicates the time.

Therefore, $\left(\bm{\beta_{t}^{(i)}}\right)_{t\in \{t_1, \ldots, T\}}$ represents the bivariate time series of period parameters estimated using the models in Section~\ref{Sec:Models}. We simultaneously model this time series using a multivariate random walk with drift:
\begin{equation}\label{eq:mrwd} 
\bm{\beta_{t}^{(i)}}=\bm{\beta_{t-1}^{(i)}}+\bm{\delta^{(i)}}+\bm{\epsilon_t^{(i)}}, \quad \bm{\epsilon_t^{(i)}} \sim N(\bm{0},\bm{\Sigma}^{(i)})
\end{equation}
where $\bm{\delta^{(i)}}$ is the vector of drift parameters that drive the dynamics for population $i$, and $\bm{\Sigma^{(i)}}$ is the $(n_{i} \times n_{i})$ variance-covariance matrix of the multivariate white noise $\bm{\epsilon_{t}^{(i)}}$.

Using the in-sample data, we compute $h$-step-ahead point forecasts and assess out-of-sample accuracy by comparing these forecasts with the actual out-of-sample data. Let $T$ denote the forecast origin and $h$ the forecast horizon. The $h$-step-ahead forecast of the period parameter is given by:
\begin{equation*}
\bm{\widehat{\beta}_{T}^{(i)}}(h)=E\left[ \bm{\beta_{T+h}^{(i)}} \Big\vert \bm{\beta_{T}^{(i)}},\bm{\beta_{t_\tau-1}^{(i)}},\dots,\bm{\beta_{1}^{(i)}}  \right],
\end{equation*}

The $h$-step-ahead forecast for the multivariate random walk with drift model, as specified in~\eqref{eq:mrwd}, can be expressed as:
\begin{equation*}
\bm{\widehat{\beta}_{T}^{(i)}}(h)=\bm{\beta_{T}^{(i)}}+\bm{\delta^{(i)}} h,
\end{equation*}

We evaluate the model's performance using the mean absolute error (MAE), root mean square error (RMSE), and mean absolute percentage error (MAPE). Let $T$ represent the final year of the forecast, and $t_\tau$ (where $t_1 < t_\tau < T$) denote the calendar year corresponding to the last observation in the \textit{in-sample} set.

We calculate the MAE, RMSE and MAPE for the differences in mortality between the two populations~$i$ as follows:
\begin{align*}
\text{MAE}&=\frac{\sum_{t = t_{\tau}+h}^{T} \mid G_{t} - \widehat{G}_{t} \mid}{(T-t_{\tau}-1)}, \\
\text{RMSE}&=\sqrt{\frac{\sum_{t = t_{\tau}+h}^{T} \left(G_{t} - \widehat{G}_{t}\right)^2}{(T-t_{\tau}-1)}}, \\ 
\text{MAPE}&=\frac{1}{(T-t_{\tau}-1)}\sum_{t = t_{\tau}+h}^{T} \frac{|G_{t} - \widehat{G}_{t}|}{G_t}\times 100, 
\end{align*}
where $G_{t}$ denotes the out-of-sample holdout data, and $\widehat{G}_t$ denotes its estimates. 

\section{Application to cancer and circulatory diseases}\label{sec:res}

We applied the three models, treating age and years as categorical variables with the first category as the reference. This approach yielded two sets of age and period effects for each model, corresponding to the $\beta$ parameters introduced in Section~\ref{Sec:Models}.

We compare the accuracy of the three proposed models in estimating the differences in death counts ($\widehat{G}_{xt}$), both in-sample and out-of-sample. Specifically, we use RMSE, MAE, and MAPE for both in-sample and out-of-sample evaluations, and further assess goodness-of-fit using BIC, AIC, and AICc. To compare the predictive accuracy of the competing forecasts, we employ the Diebold-Mariano test.

Estimating and forecasting mortality is inherently challenging, particularly when introducing a novel approach. This complexity arises from the evolving nature of mortality dynamics, which has changed significantly over the past thirty to forty years due to both endogenous and exogenous factors, such as improvements in social and health conditions \citep{rau2008continued}. Given this, each statistical model must be tested and calibrated to determine the period that best reflects the estimated parameters. In other words, it is crucial to identify the optimal period length and age range for feeding the model. These factors impact the estimation of model parameters and, consequently, the model's reliability, as measured by backtesting errors.

This aspect is frequently underestimated in mortality studies. It is important to note that this issue is not limited to our model but is relevant to all mortality models, which should be subjected to a sensitivity analysis regarding period and age structure variations. To address this, we conduct a sensitivity analysis to evaluate how performance varies with different estimation period lengths and age ranges. For our analysis, we consider two distinct age ranges:
\begin{inparaenum}
\item[(i)] complete age range, from 0 to 80+ years; and 
\item[(ii)] a specific focus on the adult age range, from 40 years onwards. For both age ranges, we examine three estimation periods: 40 years (1961-2000), 30 years (1971-2000), and 20 years (1981-2000).
\end{inparaenum}
All results are evaluated and compared over the same forecasting horizon, spanning the period 2001-2015, for consistency in the out-of-sample exercise.

\begin{table}[!htb]
\centering
\tabcolsep 0.065in
\small
\caption{\small{Accuracy evaluation of the three models in the in-sample and out-of-sample fitting over three baseline and forecasting periods for all the ages. Bold and underlined numbers represent the lowest and second lowest values among the three models for each fitting period, respectively.}\label{Tab:InOutAge0}}
\begin{tabular}{@{}lrrrrrrrrr@{}}
\toprule
\multicolumn{1}{c}{} & \multicolumn{6}{c|}{In-Sample} & \multicolumn{3}{c}{Out-of-Sample} \\ \cline{2-10} 
Model & \multicolumn{1}{l}{BIC} & \multicolumn{1}{l}{AIC} & \multicolumn{1}{l}{AICc} & RMSE & MAE & \multicolumn{1}{r|}{MAPE} & RMSE & MAE & MAPE \\ \hline
\multicolumn{10}{l}{\hspace{-.07in}{Baseline 1961-2000, Forecast 2001-2015}} \\ \hline
Skellam & \multicolumn{1}{l}{\textbf{9884.40}} & \multicolumn{1}{l}{\textbf{9377.92}} & \multicolumn{1}{l}{\textbf{9422.57}} & \textbf{466.79} & \textbf{235.49} & \multicolumn{1}{r|}{\underline{0.89}} & \textbf{1624.31} & \textbf{778.76} & \textbf{1.26} \\
Double Poisson & 299456.57 & 298950.10 & 298994.74 & 1557.18 & 813.07 & \multicolumn{1}{r|}{0.98} & 2627.33 & 1700.54 & 2.95 \\
Bivariate Poisson & \underline{292884.12} & \underline{292294.79} & \underline{292340.31} & \underline{1539.02} & \underline{791.50} & \multicolumn{1}{r|}{\textbf{0.85}} & \underline{2593.86} & \underline{1666.44} & \underline{2.49} \\ \hline
\multicolumn{10}{l}{\hspace{-.07in}{Baseline 1971-2000, Forecast 2001-2015}} \\ \hline
Skellam & \textbf{7305.39} & \textbf{6915.83} & \textbf{6956.86} & \textbf{495.35} & \textbf{238.58} & \multicolumn{1}{r|}{\textbf{0.53}} & \textbf{1475.54} & \textbf{803.09} & \textbf{0.92} \\
Double Poisson & 185354.15 & 184964.59 & 185005.62 & 1380.38 & 748.31 & \multicolumn{1}{r|}{0.76} & 2194.91 & 1413.12 & 2.05 \\
Bivariate Poisson & \underline{184891.00} & \underline{184432.74} & \underline{184474.77} & \underline{1375.31} & \underline{740.18} & \multicolumn{1}{r|}{\underline{0.71}} & \underline{2182.50} & \underline{1398.93} & \underline{1.87} \\ \hline
\multicolumn{10}{l}{\hspace{-.07in}{Baseline 1981-2000, Forecast 2001-2015}} \\ \hline
Skellam & \textbf{4765.96} & \textbf{4490.27} & \textbf{4529.64} & \textbf{520.92} & \textbf{233.36} & \multicolumn{1}{r|}{\textbf{0.29}} & \textbf{1406.80} & \textbf{865.64} & \textbf{0.42} \\
Double Poisson & 91332.70 & 91057.01 & 91096.38 & 1124.21 & 589.59 & \multicolumn{1}{r|}{0.47} & 1691.55 & 1098.81 & 1.18 \\
Bivariate Poisson & \underline{91316.54} & \underline{90986.42} & \underline{91027.04} & \underline{1122.08} & \underline{585.60} & \multicolumn{1}{r|}{\underline{0.44}} & \underline{1687.16} & \underline{1090.11} & \underline{1.04} \\ 
\bottomrule
\end{tabular}
\end{table}

The accuracy results, and the goodness-of-fit metrics related to the in-sample, are presented in Tables~\ref{Tab:InOutAge0} and~\ref{Tab:InOutAge40} for the full age range and ages 40+, respectively. The tables reveal several insights, the age-period model based on the Skellam distribution consistently provides superior in-sample estimates, demonstrating higher accuracy and superior goodness-of-fit across all periods and both age ranges.

\begin{table}[!htb]
\tabcolsep 0.04in
\centering
\small
\caption{\small{Accuracy evaluation of the three models, both in-sample and out-of-sample, across three baseline and forecasting periods for ages 40 and older. Bold and underlined numbers, respectively, represent the lowest and second lowest values among the three models for each fitting period.}}
\label{Tab:InOutAge40}
\begin{tabular}{@{}lrrrrrrrrrl@{}}
\toprule
\multicolumn{1}{c}{} & \multicolumn{6}{c|}{In-Sample} & \multicolumn{3}{c}{Out-of-Sample} &  \\ \cline{2-10}
Model & \multicolumn{1}{l}{BIC} & \multicolumn{1}{l}{AIC} & \multicolumn{1}{l}{AICc} & RMSE & MAE & \multicolumn{1}{r|}{MAPE} & RMSE & MAE & MAPE &  \\ \cline{1-10}
\multicolumn{10}{l}{\hspace{-.05in}{Baseline 1961-2000, Forecast 2001-2015, Age $\geq 40$}} &  \\ \cline{1-10}
Skellam & \textbf{9452.07} & \textbf{9079.01} & \textbf{9149.82} & \textbf{601.12} & \textbf{372.36} & \multicolumn{1}{r|}{\textbf{0.18}} & \textbf{2391.84} & \textbf{1561.41} & \textbf{0.27} &  \\
Double Poisson & 258270.67 & 257897.60 & 257968.42 & 2085.50 & 1379.95 & \multicolumn{1}{r|}{0.57} & 3534.51 & 2950.76 & 0.87 &  \\
Bivariate Poisson & \underline{ 229753.76} & \underline{ 229309.58} & \underline{ 229382.14} & \underline{ 1952.41} & \underline{ 1186.06} & \multicolumn{1}{r|}{\underline{ 0.40}} & \underline{ 3109.66} & \underline{ 2474.85} & \underline{ 0.61} &  \\ \cline{1-10}
\multicolumn{10}{l}{\hspace{-.05in}{Baseline 1971-2000, Forecast 2001-2015, Age $\geq 40$}} &  \\ \cline{1-10}
Skellam & \textbf{4439.31} & \textbf{4165.83} & \textbf{4226.47} & \textbf{633.52} & \textbf{380.14} & \multicolumn{1}{r|}{\textbf{0.17}} & \textbf{1350.22} & \textbf{890.35} & \textbf{0.20} &  \\
Double Poisson & 169287.45 & 169013.97 & 169074.61 & 1864.26 & 1280.84 & \multicolumn{1}{r|}{0.42} & 2956.41 & 2469.65 & 0.72 &  \\
Bivariate Poisson & \underline{ 156421.77} & \underline{ 156091.32} & \underline{ 156153.88} & \underline{ 1783.47} & \underline{ 1126.53} & \multicolumn{1}{r|}{\underline{ 0.34}} & \underline{ 2634.37} & \underline{ 2114.51} & \underline{ 0.53} &  \\ \cline{1-10}
\multicolumn{10}{l}{\hspace{-.05in}{Baseline 1981-2000, Forecast 2001-2015, Age $\geq 40$}} &  \\ \cline{1-10}
Skellam & \textbf{4089.25} & \textbf{3910.44} & \textbf{3962.35} & \textbf{677.45} & \textbf{408.04} & \multicolumn{1}{r|}{\textbf{0.12}} & \textbf{1156.49} & \textbf{931.80} & \textbf{0.21} &  \\
Double Poisson & 85926.79 & 85747.99 & 85799.89 & 1529.94 & 1050.06 & \multicolumn{1}{r|}{0.30} & 2327.94 & 1973.18 & 0.54 &  \\
Bivariate Poisson & \underline{ 81892.53} & \underline{ 81671.02} & \underline{ 81725.22} & \underline{ 1485.71} & \underline{ 964.05} & \multicolumn{1}{r|}{\underline{ 0.25}} & \underline{ 2183.59} & \underline{ 1771.90} & \underline{ 0.41} &  \\ 
\bottomrule
\end{tabular}
\end{table}

In terms of out-of-sample estimation, the sensitivity analysis indicates that the model based on the Skellam distribution still outperforms the other two models. All the models have better performances on the more recent windows, which is not surprising because this is the period with the most regular longevity dynamics \citep{nigri2022relationship}. 

\begin{figure}[!htb]
\centering
\includegraphics[width=0.72\linewidth]{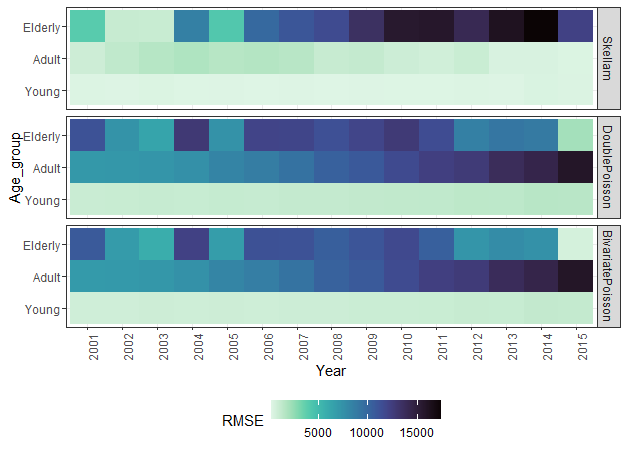}
\caption{\small{Root Mean Square Error (RMSE) of Forecasting Differences in Death Counts: Baseline 1961-2000; Ages 0-80+. Dark colours represent inferior forecasting accuracy.}}\label{Fig:2}
\end{figure}

To offer a more comprehensive analysis and enhance the understanding of the models' behavior, we present a two-phase graphical investigation. In Figure~\ref{Fig:2}, we provide heatmaps illustrating the out-of-sample RMSE across different age groups: young ages (0-24), adult ages (25-59), and elderly ages (60+). Additionally, scatterplots are presented to demonstrate the model fit, both in-sample and out-of-sample, over time for selected ages. The results for the period 1961-2000 are detailed below, with corresponding results for other periods available in Appendix~\ref{Appendix}.

\begin{figure}[!htb]
\centering
\includegraphics[width=0.72\linewidth]{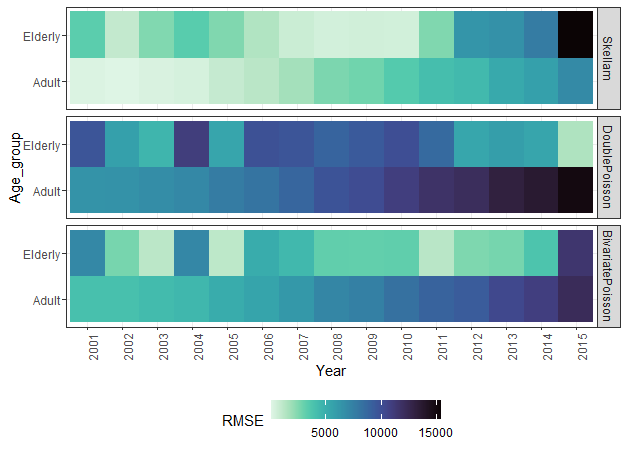}
\caption{\small{Root Mean Square Error of forecasting estimation for differences in death counts. Estimations are based on the baseline period 1961-2000 for ages 40-80+. Dark colours represent inferior forecasting accuracy.}}\label{Fig:4}
\end{figure}

Figure~\ref{Fig:2} shows that the models achieve better accuracy at younger ages, although even in this age group, the Skellam regression provides slightly more accurate forecasts. The accuracy advantage of the Skellam-based model becomes more pronounced at adult ages, as also confirmed in Figure~\ref{Fig:4}. However, among older individuals, especially as the forecasting horizon approaches the last observations, the accuracy of the bivariate Poisson and double Poisson models surpasses that of the Skellam model. The two Poisson-based models exhibit fairly similar trends, though the bivariate Poisson model displays slightly lighter colors at adult and older ages.

\begin{figure}[!htb]
\centering
\includegraphics[width=0.72\linewidth]{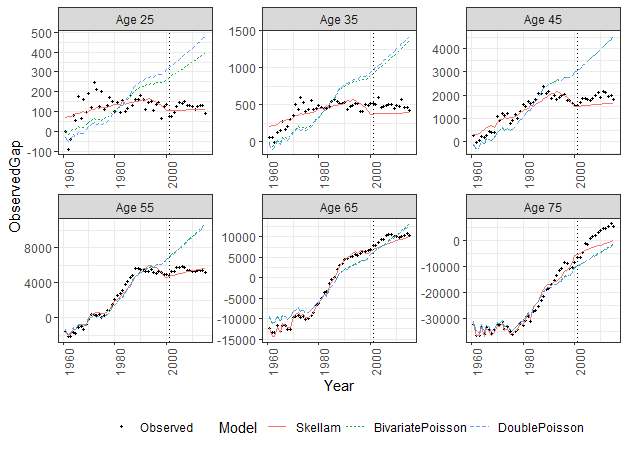}
\caption{\small{In-sample and out-of-sample analysis for specific ages, with estimations based on the baseline period 1961-2000; ages 0-80+.}}\label{Fig:3}
\end{figure}

\begin{figure}[!htb]
\centering
\includegraphics[width=\linewidth]{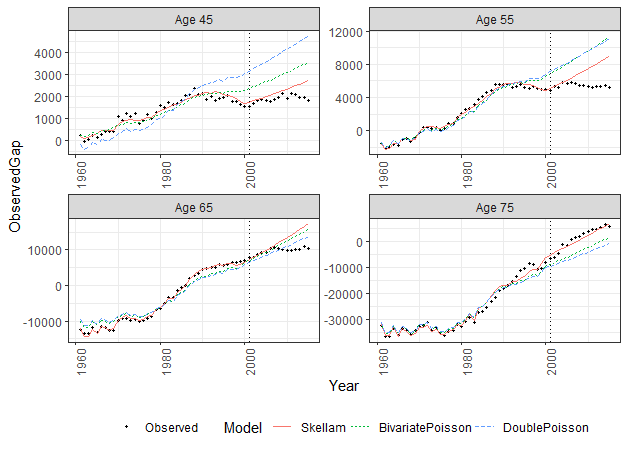}
\caption{\small{In-sample and out-of-sample exercise for specific ages. Estimations based on the baseline period 1961-2000; ages 40-80+.}} \label{Fig:5}
\end{figure}

Furthermore, we can observe from Figures~\ref{Fig:3} and~\ref{Fig:5}, both in-sample and out-of-sample results demonstrate that the Skellam-based model provides an exceptional fit to the data compared to the other models, and importantly, it maintains this consistency over time. When the model is trained on a shorter baseline period (see Appendix~\ref{Appendix_B}), the accuracy decreases but remains outstanding for some ages, mainly in the in-sample fitting.

\subsection{Diebold-Mariano (DM) test}

To compare the predictive accuracy of competing forecasts, we employ the DM test \citep{DM95}. In essence, the forecast error $e_{it}$ is defined as the difference between the values predicted by model $i$ and the actual observed values at time $t$. The loss associated with the forecast from model $i$ is assumed to be $g(e_{it}) = e_{it}^2$.
The two forecasts are considered to have equal accuracy if and only if the expected value of the loss differential, $d_t = g(e_{1t}) - g(e_{2t})$, is zero for all $t$. The DM test statistic is defined as follows:
\begin{equation}
\text{DM} = \frac{\overline{d}}{\sqrt{\frac{s}{N}}},
\end{equation}
where $\overline{d}$ is the sample mean of the loss differential, $s$ is the sample variance of the loss differential, and $N$ represents the sample size.
The null hypothesis of the test states that the competing models have equal forecast accuracy, i.e., $H_0: E[d_t] = 0, \, \forall t$, while the alternative hypothesis is $H_1: E[d_t] \neq 0, \, \forall t$. Under $H_0$, the DM statistic is asymptotically distributed as a standard normal distribution with a mean of 0 and a standard deviation of 1.

We apply the DM test to compare the forecast accuracy of the Skellam Model against the double Poisson and bivariate Poisson models. The results of the DM test are reported in Tables \ref{tab:DM1} and \ref{tab:DM2}. When the null hypothesis is rejected at a certain level of significance, it indicates that the differences between the two models are statistically significant, and the forecasting accuracy of the first model (Skellam, in our study) is superior to that of the second model. From Tables, we observe that the forecasts produced by the Skellam approach consistently exhibit higher accuracy compared to those from the other two models.

\begin{table}[!htb]
\centering
\tabcolsep 0.14in
\renewcommand{\arraystretch}{0.95}
\begin{tabular}{@{}lll|ll|ll@{}}
\cmidrule{2-7}
&\multicolumn{2}{c|}{\begin{tabular}[c]{@{}c@{}}Baseline 1961-2000\\ Forecast 2001-2015\end{tabular}} &
\multicolumn{2}{c|}{\begin{tabular}[c]{@{}c@{}}Baseline 1971-2000\\ Forecast 2001-2015\end{tabular}} &
\multicolumn{2}{c}{\begin{tabular}[c]{@{}c@{}}Baseline 1981-2000\\ Forecast 2001-2015\end{tabular}} \\ \cmidrule{2-7}
\multicolumn{1}{l|}{\textbf{\begin{tabular}[c]{@{}l@{}}Age\\ Group\end{tabular}}} &
\multicolumn{1}{c}{\textbf{\begin{tabular}[c]{@{}c@{}}Double\\ Poisson\end{tabular}}} &
\multicolumn{1}{c|}{\textbf{\begin{tabular}[c]{@{}c@{}}Bivariate\\ Poisson\end{tabular}}} &
\multicolumn{1}{c}{\textbf{\begin{tabular}[c]{@{}c@{}}Double\\ Poisson\end{tabular}}} &
\multicolumn{1}{c|}{\textbf{\begin{tabular}[c]{@{}c@{}}Bivariate\\ Poisson\end{tabular}}} &
\multicolumn{1}{c}{\textbf{\begin{tabular}[c]{@{}c@{}}Double\\ Poisson\end{tabular}}} &
\multicolumn{1}{c}{\textbf{\begin{tabular}[c]{@{}c@{}}Bivariate\\ Poisson\end{tabular}}} \\
\multicolumn{1}{l|}{\textbf{0-4}}   & $-12.014^{***}$ & $-10.612^{***}$ & $-9.687^{***}$  & $-8.694^{***}$  & $-6.527^{***}$  & $-6.09^{***}$  \\
\multicolumn{1}{l|}{\textbf{5-9}}   & $-14.374^{***}$ & $-14.457^{***}$ & $-13.877^{***}$ & $-14.092^{***}$ & $-9.473^{***}$  & $-9.251^{***}$ \\
\multicolumn{1}{l|}{\textbf{10-14}} & $-12.071^{***}$ & $-11.882^{***}$ & $-12.472^{***}$ & $-12.607^{***}$ & $-9.468^{***}$  & $-9.29^{***}$  \\
\multicolumn{1}{l|}{\textbf{15-19}} & $-10.396^{***}$ & $-9.462^{***}$  & $-10.404^{***}$ & $-10.096^{***}$ & $-8.017^{***}$  & $-7.605^{***}$ \\
\multicolumn{1}{l|}{\textbf{20-24}} & $-12.004^{***}$ & $-11.274^{***}$ & $-12.609^{***}$ & $-12.38^{***}$  & $-10.182^{***}$ & $-9.71^{***}$  \\
\multicolumn{1}{l|}{\textbf{25-29}} & $-10.572^{***}$ & $-9.715^{***}$  & $-9.752^{***}$  & $-9.224^{***}$  & $-6.643^{***}$  & $-5.955^{***}$ \\
\multicolumn{1}{l|}{\textbf{30-34}} & $-7.308^{***}$  & $-6.681^{***}$  & $-6.37^{***}$   & $-6.076^{***}$  & $-4.349^{**}$   & $-4.08^{**}$   \\
\multicolumn{1}{l|}{\textbf{35-39}} & $-7.818^{***}$  & $-7.427^{***}$  & $-6.586^{***}$  & $-6.376^{***}$  & $-4.01^{**}$    & $-3.76^{**}$   \\
\multicolumn{1}{l|}{\textbf{40-44}} & $-7.525^{***}$  & $-7.289^{***}$  & $-6.701^{***}$  & $-6.602^{***}$  & $-4.381^{**}$   & $-4.271^{**}$  \\
\multicolumn{1}{l|}{\textbf{45-49}} & $-8.928^{***}$  & $-8.692^{***}$  & $-9.031^{***}$  & $-8.953^{***}$  & $-5.694^{***}$  & $-5.613^{***}$ \\
\multicolumn{1}{l|}{\textbf{50-54}} & $-7.728^{***}$  & $-7.591^{***}$  & $-8.17^{***}$   & $-8.125^{***}$  & $-8.028^{***}$  & $-8.004^{***}$ \\
\multicolumn{1}{l|}{\textbf{55-59}} & $-5.366^{***}$  & $-5.354^{***}$  & $-5.229^{***}$  & $-5.232^{***}$  & $-4.771^{***}$  & $-4.777^{***}$ \\
\multicolumn{1}{l|}{\textbf{60-64}} & $-3.154^{**}$   & $-3.253^{**}$   & $-2.984^{*}$    & $-3.026^{**}$   & $-2.443^{*}$    & $-2.479^{*}$   \\
\multicolumn{1}{l|}{\textbf{65-69}} & $-0.575^{}$     & $-0.637^{}$     & $2.185^{*}$     & $1.95^{}$       & $5.742^{***}$   & $5.559^{***}$  \\
\multicolumn{1}{l|}{\textbf{70-74}} & $-8.478^{***}$  & $-7.14^{***}$   & $-3.183^{**}$   & $-2.876^{*}$    & $2.194^{*}$     & $2.281^{*}$    \\
\multicolumn{1}{l|}{\textbf{75-79}} & $-9.641^{***}$  & $-9.185^{***}$  & $-7.294^{***}$  & $-7.065^{***}$  & $-3.891^{**}$   & $-3.773^{**}$  \\
\multicolumn{1}{l|}{\textbf{80+}}   & $2.449^{*}$     & $2.405^{*}$     & $2.155^{*}$     & $2.121^{}$      & $0.477^{}$      & $0.419^{}$     \\ 
\bottomrule
\end{tabular}
\caption{\small The DM test between Skellam-based forecast and double Poisson and bivariate Poisson models, respectively. Evaluation period 2001–2015 for age groups from 0 to 80+. Asterisks $^{***}$,  $^{**}$ and $^{*}$ indicate two-side significance at 0.1\%, 1\% and 5\% level, respectively}\label{tab:DM1}
\end{table}

\begin{table}[!htb]
\centering
\tabcolsep 0.17in
\renewcommand{\arraystretch}{0.95}
\begin{tabular}{@{}lll|ll|ll@{}}
\cmidrule{2-7}
 &
  \multicolumn{2}{c|}{\begin{tabular}[c]{@{}c@{}}Baseline 1961-2000\\ Forecast 2001-2015\end{tabular}} &
  \multicolumn{2}{c|}{\begin{tabular}[c]{@{}c@{}}Baseline 1971-2000\\ Forecast 2001-2015\end{tabular}} &
  \multicolumn{2}{c}{\begin{tabular}[c]{@{}c@{}}Baseline 1981-2000\\ Forecast 2001-2015\end{tabular}} \\ \cmidrule{2-7} 
\multicolumn{1}{l|}{\textbf{\begin{tabular}[c]{@{}l@{}}Age\\ Group\end{tabular}}} &
  \multicolumn{1}{c}{\textbf{\begin{tabular}[c]{@{}c@{}}Double\\ Poisson\end{tabular}}} &
  \multicolumn{1}{c|}{\textbf{\begin{tabular}[c]{@{}c@{}}Bivariate\\ Poisson\end{tabular}}} &
  \multicolumn{1}{c}{\textbf{\begin{tabular}[c]{@{}c@{}}Double\\ Poisson\end{tabular}}} &
  \multicolumn{1}{c|}{\textbf{\begin{tabular}[c]{@{}c@{}}Bivariate\\ Poisson\end{tabular}}} &
  \multicolumn{1}{c}{\textbf{\begin{tabular}[c]{@{}c@{}}Double\\ Poisson\end{tabular}}} &
  \multicolumn{1}{c}{\textbf{\begin{tabular}[c]{@{}c@{}}Bivariate\\ Poisson\end{tabular}}} \\
\multicolumn{1}{l|}{\textbf{40-44}} & $-7.931^{***}$ & $-3.03^{**}$   & $-7.209^{***}$ & $-1.056^{}$    & $-4.687^{***}$ & $4.388^{**}$   \\
\multicolumn{1}{l|}{\textbf{45-49}} & $-9.848^{***}$ & $-7.663^{***}$ & $-9.217^{***}$ & $-6.603^{***}$ & $-6.003^{***}$ & $0.118^{}$     \\
\multicolumn{1}{l|}{\textbf{50-54}} & $-8.959^{***}$ & $-7.367^{***}$ & $-8.179^{***}$ & $-7.118^{***}$ & $-7.972^{***}$ & $-7.325^{***}$ \\
\multicolumn{1}{l|}{\textbf{55-59}} & $-7.52^{***}$  & $-6.215^{***}$ & $-5.752^{***}$ & $-5.276^{***}$ & $-5.41^{***}$  & $-5.083^{***}$ \\
\multicolumn{1}{l|}{\textbf{60-64}} & $-6.247^{***}$ & $-5.582^{***}$ & $-3.964^{**}$  & $-3.938^{**}$  & $-3.689^{**}$  & $-3.661^{**}$  \\
\multicolumn{1}{l|}{\textbf{65-69}} & $2.697^{*}$    & $3.057^{**}$   & $-2.187^{*}$   & $-2.405^{*}$   & $-0.782^{}$    & $-1.334^{}$    \\
\multicolumn{1}{l|}{\textbf{70-74}} & $-2.575^{*}$   & $-0.63^{}$     & $-9.03^{***}$  & $-6.402^{***}$ & $-3.997^{**}$  & $-1.568^{}$    \\
\multicolumn{1}{l|}{\textbf{75-79}} & $-8.342^{***}$ & $-7.642^{***}$ & $-8.558^{***}$ & $-7.034^{***}$ & $-4.874^{***}$ & $-3.621^{**}$  \\
\multicolumn{1}{l|}{\textbf{80+}}   & $2.576^{*}$    & $2.144^{}$     & $1.733^{}$     & $0.938^{}$     & $-0.399^{}$    & $-1.824^{}$    \\ \bottomrule
\end{tabular}
\caption{\small The DM test between Skellam-based forecast and double Poisson and bivariate Poisson models, respectively. Evaluation period 2001–2015 for age groups from 40 to 80+. Asterisks $^{***}$,  $^{**}$ and $^{*}$ indicate two-side significance at 0.1\%, 1\% and 5\% level, respectively}\label{tab:DM2}
\end{table}

\section{Conclusion}\label{sec:conc}

We propose three distinct modeling approaches to analyze death counts and their differences, each grounded in the fundamental assumption that deaths at each age and year follow a Poisson distribution. Our approaches include three age-period models: the classic double Poisson, the more advanced bivariate Poisson, and the novel Skellam distribution. These models represent a progression from traditional methods to more sophisticated and innovative techniques in studying mortality gaps. 

We applied the three modeling approaches to analyze differences in death counts for two major causes of death in Italy: cancer and diseases of the circulatory system. By focusing on causes of death, we bypass the need to account for risk exposures, as the exposure populations are consistent for both causes, allowing us to directly compare mortality rates and death counts.

Our comparison of the three models reveals that the Skellam distribution model, an innovative approach in this context, provides the best fit to the data and requires the least amount of input information.

Given the complexity of mortality, which is influenced by numerous factors, no single statistical model can fully capture its dynamics. To address this, our study includes a sensitivity analysis to evaluate how performance varies with different estimation periods and age ranges.

Leveraging model selection criteria such as AICc and BIC in the in-sample study, and also exploiting both metric errors and hypothesis tests (DM test) for the out-of-sample exercise, we found that the Skellam distribution is valuable for jointly and consistently modeling mortality differences. Its versatility also allows for generalization to other contexts, such as comparing deaths between different sexes, countries, or economic conditions. Utilizing the Skellam distribution enables the development of coherent forecasting models for these scenarios.

All codes, data and supplementary results are available on the online Repository in 
\href{https://osf.io/cn5zd/?view_only=f8f60595e25040019eca93653fd4c1b4}{OSF}.

There are at least two ways in which our work can be further extended, and we briefly outline two: 
\begin{inparaenum}
\item[1)] Leveraging the Skellam-based framework to improve the forecasting of the death count of one of the two populations, such as one of the two causes. Combining forecasts for one of the two causes using Poisson methods with gap forecasts made using the more accurate Skellam method can also enhance the forecasting accuracy for the other cause of death. 
\item[2)] Including other covariates in the models to account for the geospatial dimension in the difference in the death counts of the two populations.
\end{inparaenum}



\bibliographystyle{agsm}
\bibliography{biblio}

\newpage
\appendix

\section{Out-of-sample accuracy from forecasting based on shorter baseline periods}\label{Appendix}

\begin{figure}[!htb]
\centering
\subfloat[RMSE of forecasting estimation of differences in death counts. Estimation baseline 1971-2000; ages~0-80+]
{\includegraphics[width=8.65cm]{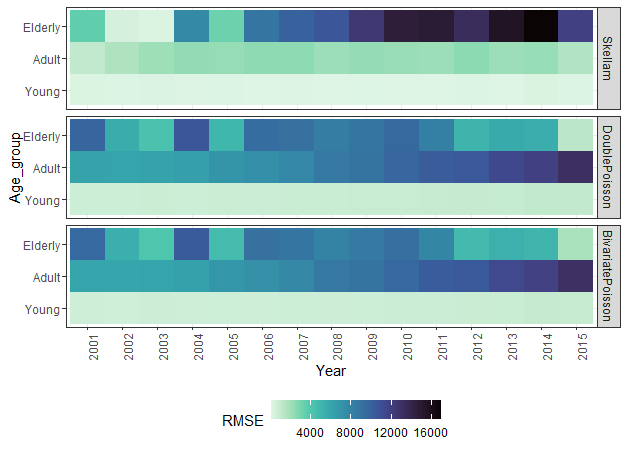}\label{Fig:RMSEbase71-AllAges}}
\quad
\subfloat[RMSE of forecasting estimation of differences in death counts. Estimation baseline 1981-2000; ages 0-80+]
{\includegraphics[width=8.65cm]{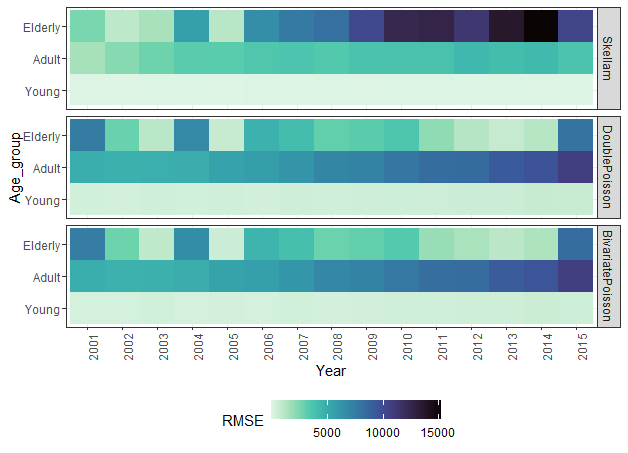}\label{Fig:RMSEbase81-AllAges}}
\\
\subfloat[RMSE of forecasting estimation of differences in death counts. Estimation baseline 1971-2000; ages 40-80+]
{\includegraphics[width=8.65cm]{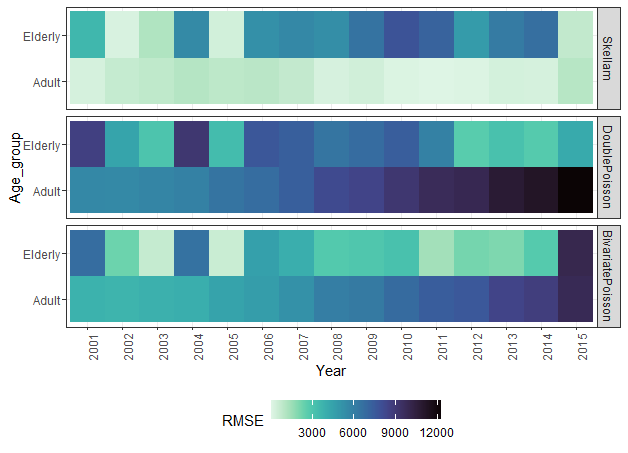}\label{Fig:RMSEbase71-40+}}
\quad
\subfloat[RMSE of forecasting estimation of differences in death counts. Estimation baseline 1981-2000; ages 40-80+]
{\includegraphics[width=8.65cm]{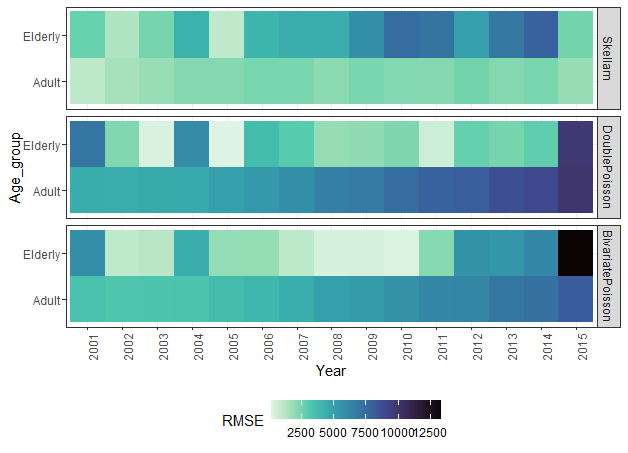}\label{Fig:RMSEbase81-40+}}
\caption{\small{Heat maps of RMSE of forecasting estimation for differences in death counts. Dark colours represent inferior forecasting accuracy.}\label{fig:appendix_A}}
\end{figure}

In Figure~\ref{fig:appendix_A}, we report an overview of the forecasting accuracy based on different baseline periods (1971-2000 and 1981-2000) and different age groups (all ages and ages over 40) for the three models.

The gap in deaths between the two causes is small at younger ages, so the accuracy of the models is better compared to older ages. No major differences between the three models are noted.

At adult ages the Skellam-based models show lighter colours than the Poisson-based models, and thus better accuracy, in Figures \ref{Fig:RMSEbase71-AllAges} and \ref{Fig:RMSEbase71-40+}, and darker colours in Figures \ref{Fig:RMSEbase81-AllAges} and \ref{Fig:RMSEbase81-40+}. This confirms that the proposed Skellam model benefits from a longer observation period to achieve remarkable accuracy in its results. At older ages with a baseline period of 30 years already, it is noticeable that the Skellam performs worse than the other two models.

\newpage

\section{In-sample and out-of-sample fittings on shorter periods}\label{Appendix_B}

\begin{figure}[!htb]
\centering
\subfloat[In-and-out-of-sample exercise for ages 0-80+. \\ Estimations baseline 1971-2000]
{\includegraphics[width=0.48\linewidth]{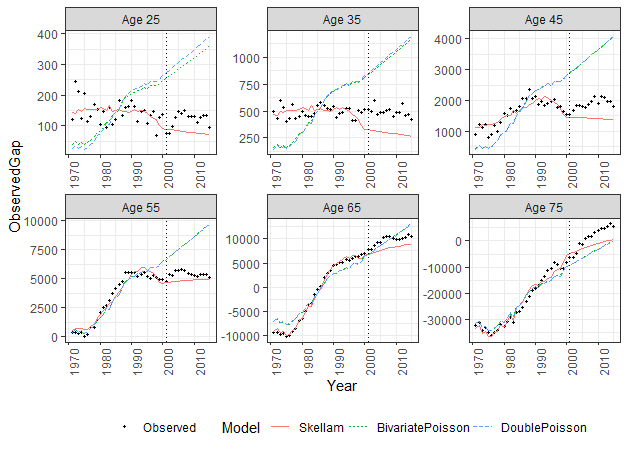}\label{Fig:In&OutFitting71-AllAges}}
\quad
\subfloat[In-and-out-of-sample exercise for ages 0-80+. \\ Estimations baseline 1981-2000]
{\includegraphics[width=0.48\linewidth]{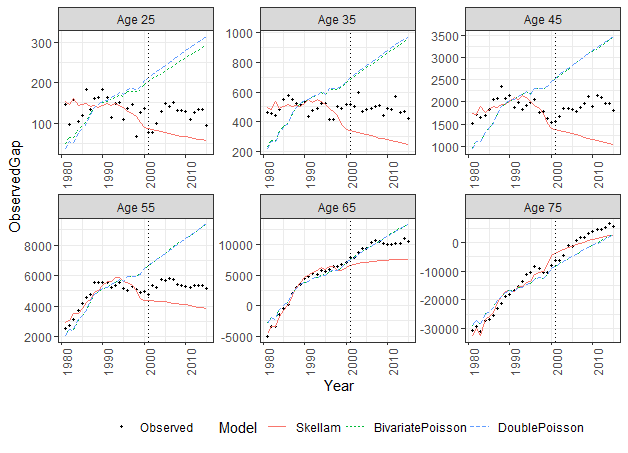}\label{Fig:In&OutFitting81-AllAges}}
\\
\subfloat[In-and-out-of-sample exercise for ages 40-80+. \\ Estimations baseline 1971-2000]
{\includegraphics[width=0.48\linewidth]{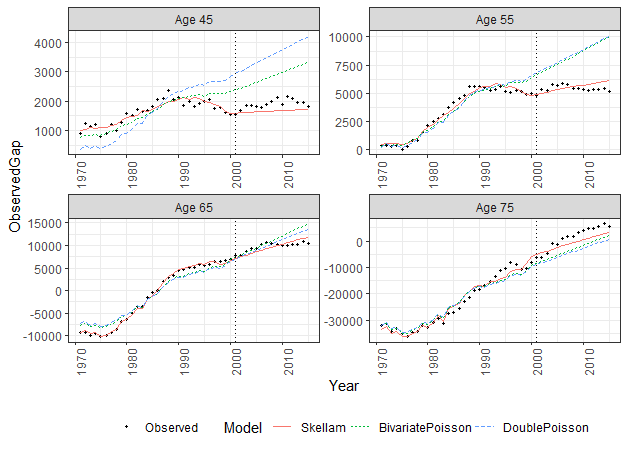}\label{Fig:In&OutFitting71-40}}
\quad
\subfloat[In-and-out-of-sample exercise for ages 40-80+. \\ Estimations baseline 1981-2000]
{\includegraphics[width=0.48\linewidth]{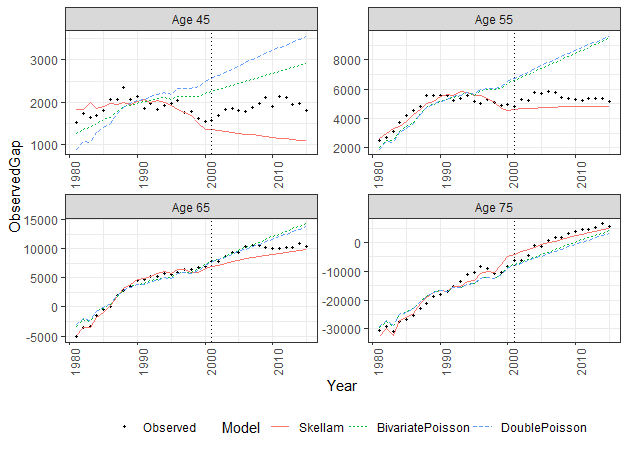}\label{Fig:In&OutFitting81-40}}
\caption{\small{Overview of all figures with different baselines and age groups.}\label{fig:appendix_B}}
\end{figure}

In Figure~\ref{fig:appendix_B}, we report an overview of the in-sample and out-of-sample fittings of the three models. In our early commentary to Figures~\ref{Fig:3} and~\ref{Fig:5} in Section~\ref{sec:res}, the Skellam-based model provides an exceptional in-sample fit to the data compared to the other models. On the other hand, in models with a very short baseline (1981-2000), the out-of-sample fitting is not exceptional. 

\end{document}